\documentstyle[aps,prb,epsfig,preprint,eqsecnum,amstex]{revtex}

\begin{document}

\newcommand{\kag}{kagom\'{e} }

\newcommand{\Kag}{KAGOM\'{E} }
\newcommand{\um}{\mu m}
\renewcommand{\eqref}[1]{Eq.(\ref{#1})}
\newcommand{\figref}[1]{Fig.\ref{#1}}
\newcommand{\HeIII}{{$^3$He~}}
\newcommand{\abs}[1]{{\left |{#1}\right |}}
\newcommand{\pipiPhi}{\frac{2\pi}{\Phi_0}}
\newcommand{\set}[1]{\{ A_{#1} \}}
\renewcommand{\vec}[1]{{\bf #1}}


\tighten


\title{Landau levels in the case of two degenerate coupled
bands: kagom\'{e} lattice tight-binding spectrum}

\author{Yi Xiao, Vincent Pelletier, Paul M. Chaikin and David A. Huse}
\address{Department of Physics, Princeton University,Princeton, NJ 08544}
\date{\today}
\maketitle

\begin{abstract}

The spectrum of charged particles hopping on a \kag lattice in a
uniform transverse magnetic field shows an unusual set of Landau
levels at low field.  They are unusual in two respects: the 
lowest Landau levels are {\it para}magnetic so their energies {\it decrease} 
linearly with increasing
field magnitude, and the spacings between the levels are not
equal.  These features are shown to
follow from the degeneracy of the energy bands in zero
magnetic field.  We give a general discussion of Landau levels in
the case of two degenerate bands, and show how the \kag lattice
tight-binding model includes one special case of this more general
problem.  We also discuss the consequences of this for the behavior
of the critical temperature of a \kag grid superconducting wire 
network, which is the experimental system that originally motivated
this work.

\end{abstract}

\pacs{PACS numbers:}


\narrowtext


The behavior of electrons in a two-dimensional periodic structure 
immersed in an external magnetic
field has been a subject of special interest over the 
past two decades ever since Hofstadter's graphical 
solution for square lattice~\cite{Hofstadter76}. 
The competition between the periodic potential
imposed by the lattice structure and the length scale 
provided by the magnetic field leads to 
frustration in the system and to intricate, detailed, and 
self-similar features for the subband 
spectrum of single-particle eigenenergies. 
The magnetic-field-induced frustration in 
two-dimensional structures has also been a 
subject of active experimental study using 
superconducting wire networks~\cite{Pannetier83,Mooij88}, as
the linearized Ginzburg-Landau equation for the superconductor and the 
Schrodinger equation for the tight-binding electron
wavefunction in these lattices share the same form~\cite{Alexander83}. 
The investigation of a variety of structures, such as the triangular 
lattice~\cite{Claro79}, the honeycomb lattice~\cite{Rammal85}, 
and recently the ${\mathcal T}_3$ 
or dice lattice~\cite{Vidal98,Pannetier01}, has revealed the 
richness and the beauty of physics in these systems. 
 
In this paper, we present our study of the Landau subband spectrum 
of the \kag lattice, 
whose distinguishing degenerate ground state has attracted extensive 
attention over the last decade.
The \kag lattice is a two-dimensional lattice of corner-sharing 
triangles (\figref{fig:kaglattice}(a)). 
The nearest-neighbor-coupled Heisenberg antiferromagnet system 
on a \kag lattice
has been shown to have an infinite number of classical ground states, 
which give rise to
a zero temperature extensive entropy~\cite{Elser89,Ritchey93}.  
However this degeneracy may be 
partially lifted by quantum or thermal 
fluctuations~\cite{Chalker92,HR92,Reimers93}. 
This so-called {\em order from disorder} mechanism~\cite{Villain80} selects the
states with the largest fluctuations (thus the highest entropy or the lowest
zero-point energy), which often are more ordered states. 

Experimentally, \kag systems have been previously studied in the context of
 adsorbed \HeIII on graphite at millikelvin temperature
~\cite{Elser89,Greywall89,Greywall90}, and the layered 
oxide $\mbox{SrCr}_{8-x}\mbox{Ga}_{4+x}\mbox{O}_{19}$    
~\cite{Ritchey93,Ramirez91}. Our recent experimental studies~\cite{Higgins00} 
of nearly perfect \kag structures using superconducting wire networks
show interesting and complex phase boundaries between the
normal and superconducting states, which 
agree well with Y.L. Lin and F. Nori's mean-field 
calculation based on quantum interference~\cite{Lin94}.
This work shows that a frustrated ground state 
exists at a magnetic field of one half of a flux quantum per
triangular plaquette. 

The lowest energy states of a charged particle in a nondegenerate
band become Landau levels when a small magnetic field $B$ is added.  The
energies of these Landau levels {\it increase} linearly in $|B|$; the
currents in these states are {\it dia}magnetic.  The
consequence of this for the critical 
temperature $T_c$ of a periodic superconducting
grid is a local maximum in $T_c$ at certain rational values of the magnetic
flux per unit cell (including, of course, zero magnetic field), 
with $T_c$ decreasing linearly in the magnitude of
the change of the field away from these special values, as has been
observed in experiments on various types of wire grids and
Josephson-junction arrays.  For the frustrated
\kag grid, on the other hand, there are two bands of lowest energy states,
and the currents in the resulting lowest Landau level of the tight-binding model 
are instead {\it para}magnetic and the state's energy {\it decreases} 
linearly in $|B|$.  This results in a {\it minimum} in
$T_c$ vs. field at the frustrating value of one-half of a flux quantum
per triangular plaquette, as is seen experimentally in the \kag
superconducting grids at this frustrating value of the applied field.

Thus from the specific \kag lattice tight-binding model, we are led to
consider the general case of Landau levels in a small magnetic field with
two degenerate bands.  In this paper we examine this more general
problem first, and then show how the
\kag lattice spectrum reduces to a special case,
which can be readily solved analytically.  We have also obtained extensive
numerical results for the \kag tight-binding spectrum, 
which agree well with our analytic results.  We conclude with a discussion
of the \kag grid superconducting wire networks that motivated this work.


\section{Two-band Landau levels}\label{sec:landau}

Here we will explore a generalization of the usual Landau levels for a
charged particle moving in two dimensions in a uniform magnetic field. 
The new case we consider is where the
particle in zero field has two bands that are degenerate at zero momentum.
This is the case in the \kag lattice tight-binding model.

What are the ``rules'' we will use to generalize the usual one-band case?
The particle is moving in two dimensions.  We look in the continuum limit of
low momentum $p$, so only look at the energy to order
$p^2$.  At this order in $p$, the spectrum of the particle's states in
zero magnetic field ($B=0$) is assumed to be isotropic in momentum space,
reflecting an underlying system that has at least three-fold rotational
symmetry.  The system is time-reversal invariant for $B=0$ and the spectrum
depends only on $|B|$, not on the sign of $B$.  (Because the motion is
two-dimensional, only the normal component of the magnetic field enters.)

For a single band, the only Hamiltonian that satisfies all these constraints
is the usual $H=\Pi^2/(2m^*)$, where 
\begin{equation}\label{eq:Pidef}
{\bf \Pi} = \vec p - \frac{q}{c} \vec A
\end{equation}
is the gauge-invariant momentum.  
In a small uniform magnetic field, this gives the
usual spectrum of equally-spaced Landau levels 
$E_n=\hbar\omega_c(n+\frac{1}{2})$, with $\omega_c=|qB|/(m^*c)$.
The currents are diamagnetic for these standard Landau levels.

If instead we have two bands that are degenerate at $\vec p =0$ for $B=0$,
then in addition to the momentum, the single-particle states are labelled
by a band index, which may be treated as a spin-1/2.  
The general Hamiltonian in
this case consists of Hermitian $2\times 2$ (Pauli) matrices operating on the
band index and Hermitian quadratic combinations of $\Pi_x$ and $\Pi_y$.
In addition to the zero-angular momentum combination $\Pi^2$ that is allowed
in the one-band case, 
there are two Hermitian operators quadratic in ${\bf\Pi}$ with
angular momentum two: $\Pi_x\Pi_y+\Pi_y\Pi_x$ and $\Pi^2_x-\Pi^2_y$, as well
as the magnetic field itself, which is proportional to the operator combination
$i(\Pi_x\Pi_y-\Pi_y\Pi_x)$. 

We have a number of constraints that restrict our choice of Hamiltonian; let us
first look at the simpler case of $B=0$.  The general Hermitian Hamiltonian that
is quadratic in $\vec p$ is  
\begin{equation}\label{eq:b=0ham}
H=\sum_{i=0}^{3}\sigma_i E_i(\vec p)~,
\end{equation}
where the $\sigma_i$ are the Pauli matrices, 
with $\sigma_0$ the identity matrix,
and the $E_i$ are real quadratic functions of $\vec p$.  
This is readily diagonalized,
with the resulting eigenenergies  
\begin{equation}\label{eq:b=0spectr}
E_{\pm}(\vec p)=E_0\pm\sqrt{E_1^2+E_2^2+E_3^2}~.
\end{equation}
Our restriction that the $B=0$ spectrum is rotationally invariant then dictates
that $E_0(\vec p)=d_0p^2$ and
\begin{equation}\label{eq:e123}
E_1^2+E_2^2+E_3^2=d_2^2p^4~, 
\end{equation}
with $d_0$, $d_2$ real constants.  Another way of writing the Hamiltonian
which explicitly breaks it into its parts with orbital angular momentum quantum
numbers $m_o=0, \pm 2$ is  
\begin{equation}\label{eq:b=0hamxyz}
H=d_0\sigma_0p^2+\vec z\cdot{\bf\sigma}p^2
+(\vec x-i\vec y)\cdot{\bf\sigma}(p_x+ip_y)^2
+(\vec x+i\vec y)\cdot{\bf\sigma}(p_x-ip_y)^2~,
\end{equation}
where $\vec x$, $\vec y$ and $\vec z$ are real 3-vectors in the space
spanned by the 3 Pauli matrices, $\sigma_1, \sigma_2, \sigma_3$.  
\eqref{eq:e123} requires that these three vectors form a mutually orthogonal
triad, with $\vec x$ and $\vec y$ being of equal magnitude.  Thus we may choose
the basis for the band index so that this triad $\vec x$, $\vec y$ and $\vec z$
are parallel to the 1, 2 and 3 axes, respectively.  This leaves a three real
parameter family of $B=0$ two-band Hamiltonians:  
\begin{equation}\label{eq:b=0ham013}
H=d_0\sigma_0p^2+d_3\sigma_3p^2
+d_1(\sigma_-(p_x+ip_y)^2+\sigma_+(p_x-ip_y)^2)~,
\end{equation}
where $\sigma_{\pm}=(\sigma_1\pm i\sigma_2)/2$ are the usual raising
and lowering operators.  The spectrum is rotationally invariant, as we required:
\begin{equation}\label{eq:b=0spectrc}
E_{\pm}(\vec p)=p^2(d_0\pm\sqrt{d_1^2+d_3^2})~.
\end{equation}
An important observation about this Hamiltonian \eqref{eq:b=0ham013}
is that it conserves a total angular momentum, provided that we define
the particles in the
bands to have an ``internal'' angular momentum 
whose quantum number $m_i=\pm 1$ is the
eigenvalue of $\sigma_3$ in this basis.  [We will see below that such
an ``internal'' angular momentum $m_i=\pm 1$ 
within each unit cell does indeed exist for
the low-energy, low-momentum states of the \kag lattice tight-binding
model.]  The operators $\sigma_{\pm}$ 
then raise/lower the internal angular momentum by 2 quanta, while
$(p_x\pm ip_y)^2$ raise/lower the orbital angular momentum by
the same amount.  Note that the operators $(p_x\pm ip_y)$ are {\it not}
simply the usual orbital angular momentum raising and lowering operators
$L_{\pm}$; in addition to changing the orbital angular moentum they also
change the radial wavefunctions.
If neither band has 
negative energy states at $B=0$, so the ground state is at $E=0$,
this corresponds to $d_0 \geq \sqrt{d_1^2+d_3^2} \geq 0$.

Now when we add the uniform magnetic field, the momentum operator $\vec p$
becomes the gauge-invariant combination $\vec\Pi = \vec p - \frac{q}{c} \vec A$.
For $B\neq 0$ the two components of $\vec\Pi$ are noncommuting variables with
a $c$-number commutator proportional to the magnetic field:
\begin{equation}\label{eq:comm}
[\Pi_x,\Pi_y]=i\frac{qB\hbar}{c}~. 
\end{equation} 
Thus we can make
harmonic-oscillator-type raising and lowering operators $a^{\dagger}$ and $a$
such that
\begin{equation}\label{eq:pix}
\Pi_x=\sqrt{\frac{|qB|\hbar}{2c}}(a^{\dagger}+a)
\end{equation}
and
\begin{equation}\label{eq:piy}
\Pi_y=i  {\rm sign}(qB)\sqrt{\frac{|qB|\hbar}{2c}}(a^{\dagger}-a)~.
\end{equation}
Since the magnetic field itself is quadratic 
in $\vec\Pi$ (\eqref{eq:comm}), a term
linear in $B$ may also be added to the Hamiltonian.  
This term may not contain the 
identity Pauli matrix, or it will generate a spectrum that 
is not invariant under
inverting $B$, but it is otherwise unrestricted.  Thus we end up with the
general two-band Hamiltonian satisfying all our ``rules'':  
\begin{equation}\label{eq:hampi}
H=(d_0\sigma_0+d_3\sigma_3)\Pi^2+d_1(\sigma_-(\Pi_x+i\Pi_y)^2
+\sigma_+(\Pi_x-i\Pi_y)^2)+B\vec b\cdot\sigma~,
\end{equation}
where $\vec b$ is a real 3-vector.

At this point let us note that the current density 
operator for this Hamiltonian
\begin{equation}\label{eq:j}
\vec J=-c\frac{\partial H}{\partial\vec A}
\end{equation} 
contains, in additional to conventional-looking terms,
a term proportional to $d_1$ that is off-diagonal in the band
index and linear in the gauge-invariant momentum $\vec \Pi$.
It is this term that allows Landau levels involving both bands
to exhibit paramagnetic currents.  These currents arise from an
``interference'' between the amplitude in one band and the gradient
of the amplitude in the other.

The transformation between the $\Pi$ operators and the raising and lowering
operators $a$ and $a^{\dagger}$ is singular at $B=0$, so it is simplest to just
consider one sign of $qB$ at a time.  In fact it is for $qB<0$ that the raising
and lowering operators $a$ and $a^{\dagger}$ also raise and lower 
the orbital angular
momentum, so let's do that case.  Since we are motivated by superconductors
we will take $B>0$ and $q=-e^*=-2e<0$.  This yields
\begin{equation}\label{eq:hamb}
H=B[\frac{\hbar e^*}{c}(d_0\sigma_0+d_3\sigma_3)(2a^{\dagger}a+1)
+\frac{2\hbar e^*d_1}{c}(\sigma_-(a^{\dagger})^2+\sigma_+a^2)
+\vec b\cdot\sigma]~.
\end{equation}
When the 1 or 2 components of $\vec b$ are nonzero, this introduces terms
that flip the band index without changing the orbital 
angular momentum, so do not
conserve total angular momentum.  We have not found a closed-form solution for
this general case, but for the \kag lattice problem we are considering, these
terms are not present.  In this case, $b_1=b_2=0$, the total angular momentum is
conserved.  The orbital angular momentum quantum number $m_o$ is also the
eigenvalue of the number operator $a^{\dagger}a$ and is restricted to be 
nonnegative for the case $qB<0$ that we are now considering.  Thus for total
angular momentum $m$ of 0 or $-1$, there is just one state, with internal
angular momentum quantum number $m_i=-1$ and $m_o=m+1$.  For $m_i+m_o=m>0$,
on the other hand, 
there are two states, with $m_i=\pm 1$.  
If $b_1=b_2=0$ so that $m$ is conserved,
the Hamiltonian then reduces to a simple 2$\times$2 matrix.  
The resulting Landau
levels are in general not equally spaced in energy, and even when there are
no negative energy states at $B=0$, some of the Landau levels may be at negative
energy for $B\neq 0$, and thus paramagnetic.  
Now will look at the \kag tight-binding model, and show
that in the low-momentum limit it realizes this two-band Hamiltonian 
in the simple case of $d_0=d_1>0$, $d_3=0$,
and $\vec b=0$.

\section{\Kag tight-binding Spectrum in Zero Field}\label{sec:zerofield}


The \kag lattice is a two dimensional periodic array 
of corner-sharing triangles with three sites per unit cell. 
As illustrated in  \figref{fig:kaglattice}(a), 
$a$ is the triangle edge length, $A$,$B$ and $C$ denote 
the three sites in a unit cell
.  The unit cell can be taken as
a rhombus of side $2a$ with angles $\frac{2\pi}3$ and $\frac {\pi}3$ at its
vertices. In the reciprocal space $\vec k=(k_x,k_y)$, 
the first Brillouin zone is a hexagon 
with a side length of $\frac {2\pi}{3a}$.
Some points on the edge of the first Brillouin zone are 
shown in \figref{fig:kaglattice}(b):   
$F(\frac {\pi}{\sqrt 3 a},0)$ and $G(\frac {\pi}{\sqrt 3 a},\frac {\pi}{3a})$, 
and  
$H(0, \frac {2\pi}{3a})$. 


We first consider the tight-binding model for a 
particle making nearest-neighbor hops on the 
\kag lattice in zero magnetic field.  The hopping matrix element is $t>0$. 
This problem is readily diagonalized.  
Since there are three sites per unit cell,
there are three bands of simultaneous eigenstates of the tight-binding
Hamiltonian and the crystal momentum:
\begin{equation}\label{eq:KagomeDispersion23}
\begin{split}
\epsilon_1(\vec k) &=\epsilon_0 - 2t,\\
\epsilon_{2,3}(\vec k) &= \epsilon_0 + t \pm t \left [1 + 
            4\cos^2(k_ya)+ 4\cos(k_ya)\cos(\sqrt 3 k_xa) \right ]^{1/2}, 
\end{split}
\end{equation}
where $\epsilon_0$ is the on-site energy of the orbitals.


For simplicity, let us set $t=1$ and $\epsilon_0=0$. 
The spectrum of the second two bands ($\epsilon_2$, $\epsilon_3$) 
ranges over the interval 
$[-2,+4]$ and the edges of this interval are reached for $\vec k=0$. 
Close to $\vec k=0$,
 these two bands are parabolic with an
isotropic effective mass.  This effective mass is negative
$m^*=-\hbar^2/(2ta^2)$ at the top of the upper band at $\epsilon=4$.
These two dispersive bands also touch at ``Dirac points'' at all
the corners of the first Brillouin zone
(such as $G$ and $H$ in \figref{fig:kaglattice}(b)), 
at energy $\epsilon_{2}=\epsilon_{3}=1$.

The first band is completely non-dispersive at energy $\epsilon_1=-2$.
Thus the ground state of this particle is highly degenerate.
Note also that the bottom of the lower dispersive band is also degenerate
with this flat band at $\vec k=0$.  Thus here at $\vec k=0$ and 
$\epsilon = -2$ we have two degenerate bands of precisely the type discussed
in the previous section of this paper.  Since one band is flat, we have
$d_0=\sqrt{d_1^2+d_3^2}$, in the notation of the 
previous section.  The dispersive band,
to quadratic order in the momentum, has energy
$\epsilon (\vec k)-\epsilon (\vec 0)=tk^2a^2$, so we have
$d_0=\sqrt{d_1^2+d_3^2}=ta^2/(2\hbar^2)$.






Some of the degenerate ground states with $\epsilon=-2$ in zero magnetic field
are shown in
 \figref{fig:kagomeMode}.  \figref{fig:kagomeMode}(a) and (b) show
the states at the Brillouin zone corners, while
\figref{fig:kagomeMode}(c) and (d) show the two degenerate states at the
zone center with $\vec k=0$.  For small magnetic field, the lowest Landau
levels are composed of linear combinations of smooth envelopes times these 
latter two $\vec k=0$ states.  The two states shown are the eigenstates of
the band operator $\sigma_3$.  By looking at their wavefunctions within a
unit cell (one triangle) we can see that $\sigma_3$ does indeed measure an
``internal'' angular momentum that has the two possible values
$m_i=\pm 1$ in this case, as anticipated above.

We can further anticipate the results of applying a magnetic field
by considering just these states. In the absence of coupling between the
bands (i.e., for $d_1=0$), the
states at the bottom of the bands are Landau quantized in the
standard fashion, and their
energy thus increases linearly with $|B|$.  On the other hand, 
the degeneracy between the bands is
split by the matrix element of the
perturbation ($d_1$) that couples them. In the present case this
perturbation is also linearly proportional to $|B|$. 
Thus whether the lowest-energy states
increase or decrease in energy depends on the relative magnitudes
of the coupling matrix element and the effective band masses. In
terms of currents, a charged particle (here, a Cooper pair)
in a given one-band Landau level has a
quantized {\it diamagnetic} dipole moment that is
independent of field.  Its energy in a field thus increases with $|B|$. 
For this specific \kag lattice case, the
wavefunctions {\it within} one unit cell also have currents and dipole
moments as in figure 2 c and d.  When these degenerate states are
coupled by the perturbation they are admixed and this can produce
{\it paramagnetic} currents.  The sum of this and the Landau diamagnetic
current can be either diamagnetic or paramagnetic for the lowest-energy 
state depending on the relative strength of the parameters.
  In this
case we find net paramagnetism in the ground states.

\section{Spectrum in a Magnetic Field}\label{sec:lowfield}

In the presence of a magnetic field $B$, 
the hopping term of the tight-binding Hamiltonian is modified 
by phase factors from the vector potential $\vec A$,~\cite{Peierls33}, 
i.e., $t \rightarrow te^{i\gamma_{ij}}$, where $\gamma_{ij}$ is 
the phase factor between sites $i$ and $j$:
\begin{equation}\label{eq:phasefactor}
\gamma_{i j}=\frac{2 \pi}{\Phi_0}\int_i^j \vec A \cdot d \vec l
~,
\end{equation} 
where $\Phi_0=hc/e^*$ is the flux quantum.
 
The (tight-binding) Schr\"{o}dinger 
equation in a magnetic field is thus the finite-difference equation:
\begin{equation}\label{eq:magneticTB}
\epsilon \phi_i=t \sum_{j}e^{i \gamma_{i j}}\phi_j, \qquad 
\end{equation}
where the sum is over all the nearest neighbors of site $i$.
It is convenient to measure the magnetic field in units of the
flux quantum per elementary triangular plaquette of the \kag
lattice.  The flux through one triangle is $\Phi=\frac {\sqrt 3}4 B a^2$.
Thus the ``filling ratio'', namely the fraction of a flux 
quantum through each triangle is
\begin{equation}  
f =\frac\Phi{\Phi_0}=\frac {\sqrt 3}4 \frac {Be^*a^2}{hc}~.
\end{equation}
The flux through an elementary hexagon is six times this, $6f\Phi_0$,
while the total flux through a unit cell of the \kag lattice is $8f\Phi_0$.

When $f$ is a rational number that can be written as $8f=k/n$, where
$k$ and $n$ are integers with no common factors, the total flux 
through $n$ unit cells of the \kag lattice is an integer multiple
of $\Phi_0$.  Then with this magnetic field, a gauge can be chosen so
that the Hamiltonian has a discrete translational invariance with a
unit cell containing $n$ unit cells of the \kag lattice, and thus
$3n$ sites.  The spectrum of the Hamiltonian then consists of $3n$
bands.  These can be obtained numerically for any such rational $f$
(for $n$ not too large); an implementation of this is 
detailed in Yi Xiao's Ph.D. thesis.~\cite{yi}  The resulting spectrum,
namely the \kag version of ``Hofstadter's butterfly'', is shown in
\figref{fig:kagomespectrum0to1half}.

The spectrum has various symmetries:  It is only the flux modulo the
flux quantum that affects the spectrum, so if $f$ is changed to $f+j$
with $j$ any integer, the spectrum is unchanged.  The spectrum is also unchanged
on changing $f$ to $-f$, because if $\psi$ is an eigenstate with energy
$\epsilon$ for field $f$, then its complex conjugate $\psi^*$ is an
eigenstate with the same energy $\epsilon$ for field $-f$.  These two symmetries
are not special to the \kag lattice.  The third symmetry is that if
$f$ is changed to $f+\frac{1}{2}$, this is the same as changing $t$ to
$-t$, so the spectrum is inverted.  Thus the highest energy states for
field near $f=\frac{1}{2}$ are equivalent to the lowest energy states
near zero field.  We will look at the behavior near zero field, but
because of this last symmetry, what we find also applies near $f=\frac{1}{2}$.

\section{Landau levels in the \Kag lattice near zero field}\label{sec:main}


For small magnetic field, the spectrum of the \kag lattice tight-binding
Hamiltonian shows standard 
equally-spaced Landau levels near the 
top of the bands, as shown in \figref{fig:UpperLandauLevel}.  Here there is just
the one band in zero field, with negative effective mass $m^*=-\hbar^2/(2ta^2)$.
The resulting Landau levels, to linear order in the field, 
should therefore have energy
\begin{equation}\label{eq:LandauLevelnear4}
\epsilon=t(4-\frac{16\pi}{\sqrt{3}}|f|(n+\frac 12))~.
\end{equation}
The numerical results for this portion of the spectrum are shown
in \figref{fig:UpperLandauLevel} for $f\geq \frac{1}{120}$.  The first few
Landau levels are clearly seen, and the asymptotic slopes at small
$f$ given by \eqref{eq:LandauLevelnear4} are shown for comparison
for the first 5 Landau levels.  At these values of $f$ the fit
is not wonderful, but it does seem to be improving with decreasing $f$,
as it should.

More interesting is the bottom of the zero-field spectrum, 
where there are two degenerate bands.
This results in non-standard Landau levels when the
field is turned on, as discussed above.  An
expanded view of this part of the spectrum is 
shown in \figref{fig:spectrumLowerLandau}.  It shows some Landau levels
whose energies are increasing linearly with field, as is normal, but
these levels are not equally-spaced.  In addition there are many states whose
energies decrease as the field is increased.  These latter states are the 
unusual feature we focus on here.  Note that because of the symmetry of
the spectrum discussed above, the same features appear at the top of
the spectrum near the frustrating value of the magnetic field of $f=1/2$.

In the one-band case, knowing the effective mass in zero field and
the charge is enough to determine the Landau level spectrum to linear
order in the magnetic field.  For the two-band case, on the other hand,
(eq. 1.11) there is possibly the $B\vec b\cdot\vec\sigma$ term, which 
vanishes at zero field, so to determine the values of the three parameters
making up $\vec b$ for the \kag tight-binding model, we cannot rely only
on our knowledge of the zero-field spectrum.  So we will now treat the
low-momentum behavior to lowest order in the field.  What we will find is
that $\vec b = 0$ and $d_3=0$.

The unit cell of the \kag lattice contains 3 sites around a triangle.
We choose a set of 3 basis states within this unit cell that
respect as much as possible the three-fold rotational symmetry of this
system about the center of the triangle.  The wavefunction of a
state is a complex number defined at each lattice point, but only
its gauge-invariant properties are physical.  These are the magnitude
of the wavefunction, and the {\it gauge-invariant phase differences} between
adjacent lattice points.  For any wavefunction defined
at the corners of the triangle in one unit cell, the total guage-invariant
phase difference added up stepping the three steps around the triangle
is $2\pi(f+m_i)$, where $m_i$ is an integer.  Note $m_i$ measures the
state's ``internal'' angular momentum.  The gauge-invariant phase
difference between two adjacent lattice points is only well-defined 
modulo $2\pi$, so let us restrict it to lie in the interval $[-\pi,\pi]$.
Thus the three of these phase differences added up must be in the
range $[-3\pi,3\pi]$, and for the small fields $f$ we consider here,
this means the total guage-invariant phase difference around the triangle
has to be near one of the three values: $-2\pi$, 0, or $2\pi$ and 
$m_i$ is either -1, 0 or 1.  The basis states we will use are those
where the guage-invariant phase differences are the same along all
edges of the triangle, being $2\pi(f+m_i)/3$, while the 
amplitudes are identical at all three points.  
In the limit of small momentum and
zero field, the upper band near energy $4t$ 
is made out of the $m_i=0$ states
while the lower bands near energy $-2t$ 
are made out of the two $m_i=\pm 1$ states.

We take the tight-binding Hamiltonian and rewrite it in terms
of these new basis states.  This step is done in a specific Landau
gauge.  We then take the low-field, low-momentum limit, assuming that
the (complex scalar) amplitudes of each of 
the basis states vary little between adjacent
unit cells, so a gradient expansion can be made.  The resulting three-band
Hamiltonian can be written in gauge-invariant form as the matrix
\begin{equation}\label{eq:secular4G}
H= t\left (
\begin{array}{ccc}
-2+\Pi^2+\frac{B}{2}    & -i\sqrt{3}(\Pi_x-i\Pi_y) & (\Pi_x-i\Pi_y)^2        \\
i\sqrt{3}(\Pi_x+i\Pi_y) & 4-2\Pi^2                 & i\sqrt{3}(\Pi_x-i\Pi_y) \\
(\Pi_x+i\Pi_y)^2        & -i\sqrt{3}(\Pi_x+i\Pi_y) &  -2+\Pi^2-\frac{B}{2}   \\
\end{array}
\right )
~,
\end{equation}
in units where $a=\hbar=1$ and $\Phi_0=2\pi$ and where the rows/columns
refer to $m_i=+1,0,-1$ in that order.  Here we have gone to the
orders in $\Pi$ and $B$ that contribute to the spectrum to linear order
in $|B|$ and quadratic order in $\Pi$.  Note that this continuum three-band
Hamiltonian conserves total angular momentum.
To reduce this to the two-band Hamiltonian for the lowest
bands only, we treat the coupling to the upper band in second-order
perturbation theory.  This eliminates the terms linear in $B$ in the
diagonal operators in the above matrix, and the two band Hamiltonian
is precisely of the form expected (1.11) with $\vec b = 0$, $d_3=0$ and
$d_0=d_1=ta^2/(2\hbar^2)$.  Finally, the Hamiltonian to this order
is simply
\begin{equation}\label{eq:hamk}
H=t[-2+\frac{4\pi f}{\sqrt{3}}[\sigma_0(2a^{\dagger}a+1)
+2(\sigma_-(a^{\dagger})^2+\sigma_+a^2)]
\end{equation}
for $f>0$.  This Hamiltonian conserves total angular momentum $m=m_o+m_i$.
The resulting Landau level spectrum to linear order in the magnetic field is
\begin{equation}\label{eq:LandauLevelnear-2}
\epsilon_{\pm}=t(-2+\frac{8\pi}{\sqrt{3}}|f|(m+\frac 12 \pm \sqrt{m^2+m+1}))~.
\end{equation}
For $m=-1$ and $m=0$ there is only the one state with $m_i=-1$, because
$m_o$ cannot be negative.  (There are no states with $m<-1$.)  
These states' energies are given by the plus sign
in the above equation; they are the two lowest 
standard one-band Landau levels made
out of only the $m_i=-1$ band.  Since they are standard one-band Landau
levels, they show diamagnetic currents.  For $m>0$, on the other hand, both
bands enter and the eigenstates are linear combinations of the
$(m+1)$th Landau level in the $m_i=-1$ band and the $(m-1)$th Landau
level in the $m_i=1$ band.  The interaction between the two bands
shifts the energies strongly, with the lower energy state being pushed
to below energy $-2t$.  The lowest energy state of all is the lower
energy (minus sign in the above equation for $\epsilon_{\pm}$) state
with $m=1$, which is at energy
\begin{equation}\label{eq:lowest}
\epsilon_{min}=-t(2+8\pi|f|(1-\frac{\sqrt{3}}{2})) \cong -t(2+3.37|f|)~.
\end{equation}
This state is a linear combination of the lowest Landau level ($m_o=0$)
in the $m_i=1$ band and the third Landau level ($m_o=2$) in the $m_i=-1$
band.  And it is the interference between these two bands that allows this
state to be paramagnetic, so its energy decreases with increasing $|B|$.




The straight lines in \figref{fig:spectrumLowerLandau} correspond to the 
lowest few of the upper (+ in Eq. 4.4) set of Landau levels.  As $f$ 
is decreased, the numerically obtained bands fit well to these expected
Landau levels.  A more detailed plot of the lower bands below $\epsilon=-2t$ 
is shown in \figref{fig:LowSpectrum}, in which
this range of the spectrum is calculated for filling ratio 
from $f=1/16$ down to $f=1/256$.  For a rational
filling ratio $f=1/8n$, where $n$ is an integer, these lower 
bands consist of $n-1$ subbands.  Counting the states, we find that
in the limit of small magnetic field
1/3 of all the states are in this set of bands below $\epsilon=-2t$,
corresponding to the number of states in the flat $\epsilon=-2t$ band
at zero field.  The lowest few Landau levels are also indicated by
the straight lines in \figref{fig:LowSpectrum}, and again the numerically
obtained bands fit well to these as $f$ is reduced.


To examine the very lowest Landau level in more detail, 
we carried out further numerical calculation of the bottom edge of the spectrum
to $f$ as low as $1/840$.  These results are shown 
in \figref{fig:LogLowEdge} 
on log-log scales. 
The straight line corresponds to the small $f$ behavior, 
$\Delta\epsilon=3.37f$, derived above. 
As can be seen, it gives an excellent fit to the numerical results in
this low field limit.

This behavior in the \kag lattice 
is certainly different from the low field Landau 
levels of most other periodic lattices
~\cite{Hofstadter76,Claro79,Rammal85,Vidal98}, where the ground
state energy increases as the magnetic field is turned on. In the case 
of the \kag latice, the presence of two degenerate 
bands in zero field allows these
unusual Landau levels to occur.

\section{Application to Superconducting Wire Networks}\label{sec:application}

The network equation for the mean-field phase boundary 
of a superconducting wire network, 
derived from linearized Ginzburg-Landau equation,  
has the same form as that of the tight-binding Schrodinger
equation in the same geometry~\cite{Alexander83}. 
For a periodic network with
equal bond length $a$, the superconducting network equation be can expressed as
\begin{equation}\label{eq:network4EqualL}
z\cos\frac{a}{\xi}\cdot\psi_i = \sum_{j} \psi_j {e^{-i\gamma_{ij}}}
\end{equation}
where $\psi_i$ is the order parameter at site $i$, 
$\xi$ is the superconducting coherence length, 
$z$ is the coordination number of the lattice,
and the sum is over all nearest-neighbor sites of $i$. 
Compared to the tight binding 
equation of \eqref{eq:magneticTB}, the correspondence 
between the two system is given by
\begin{equation}
\frac{\epsilon}{t} \Longleftrightarrow   z\cos\frac{a}{\xi}
~.
\end{equation}
The superconducting transition temperature of the network 
in a magnetic field is given
by the upper edge (for $t>0$) of the tight-binding spectrum.
For the \kag lattice, $z=4$.  The upper edge of the \kag
tight-binding spectrum is plotted in \figref{fig:upperedge}.  


For the phase boundary at the lower field limit (small $f$), 
our previous discussion of the low field
Landau levels near $\epsilon=4$ implies a linear relationship 
between the transition temperature and the
magnetic field
\begin{equation}
\Delta T_c(f)\ \sim (\frac{a}{\xi})^2 \sim |f|,
\end{equation} 
where $\Delta T_c(f)=T_c(0)-T_c(f)$ is the suppression of the 
critical temperature from
 its
zero field value.

As shown in \figref{fig:upperedge}, strong maxima or 
minima in the superconducting phase boundary appear
at magnetic fields $f=p/8$ for each integer $p$, 
which correspond to $p$ magnetic
flux quanta through each unit cell.  At each of these fields,
the tight-binding model is equivalent to a model in zero magnetic field
with the hopping matrix elements within each unit cell being complex.
Since there are three sites per unit cell, in each case there are
three bands in the spectrum.  When the magnetic field is changed slightly
to $f=\frac{p}{8} + \Delta f$, 
the small added magnetic field (of either sign) splits these
bands into Landau levels, just like near $f=0$.  For all of these fields,
except the special case $f=1/2$, the upper edge of the spectrum is
a state at crystal momentum zero in a nondegenerate band.  Thus changes
in the magnetic field produce conventional Landau levels like near $f=0$.
This causes the upper edge of the spectrum to decrease linearly in $|\Delta f|$,
resulting in a local {\it maximum}
in the superconductor's transition temperature vs. field
at those fields. 
The appearance of a local {\it minimum}
in $T_c$ at $f=1/2$ is due to the special degeneracy in the
highest band that occurs due to the
magnetic frustration at that field, resulting in Landau levels whose
energy moves linearly {\it away} from the band center 
as the field is changed from
this special value, as discussed above.

A local minimum in $T_c$ vs. field also occurs at $f=1/2$ in the
so-called ${\mathcal T}_3$ 
or dice lattice~\cite{Vidal98,Pannetier01}, but for quite
different reasons from the \kag lattice case.  The dice lattice is
the dual to the \kag lattice.  Its elementary plaquettes are
all rhombuses, and $f$ is the number of flux quanta passing
through each such rhombic plaquette.  Like the \kag lattice, the dice lattice
has three sites per unit cell.  At $f=1/2$ in
this dice lattice, the spectrum consists of three dispersionless
(infinite mass) bands~\cite{Vidal98}, with no degeneracy between the
bands.  For the dice lattice, the linear dependence of the upper edge
of the spectrum on field near $f=1/2$ arises not
from the formation of Landau levels, but from an ``orbital Zeeman effect''.
States with nonzero net magnetic moments 
of either sign and localized to just a few lattice sites
can be formed out of the highest
energy states at $f=1/2$.  The magnetic moments are due to currents 
circulating around the plaquettes and cause these states to have
energies that depend linearly on the magnetic field.  
Although the \kag latice spectrum also
has such a dispersionless band at $f=1/2$, the localized states that
can be made from this band in the \kag case
have no net magnetic moment and
thus no such orbital Zeeman effect.





We thank Duncan Haldane, Mark J. Higgins, Shobo Bhattacharya, 
Kyungwha Park, Marc Schreiber, Yeong-Lieh Lin and Franco Nori 
for helpful discussions.   
This work was supported by NSF Grants No. DMR 98-09483
,
98-02468, 99-76576 and 02-13706.

%



\begin{center}

\begin{figure}[h]
\centerline{
\begin{tabular}{cc}
\psfig{figure=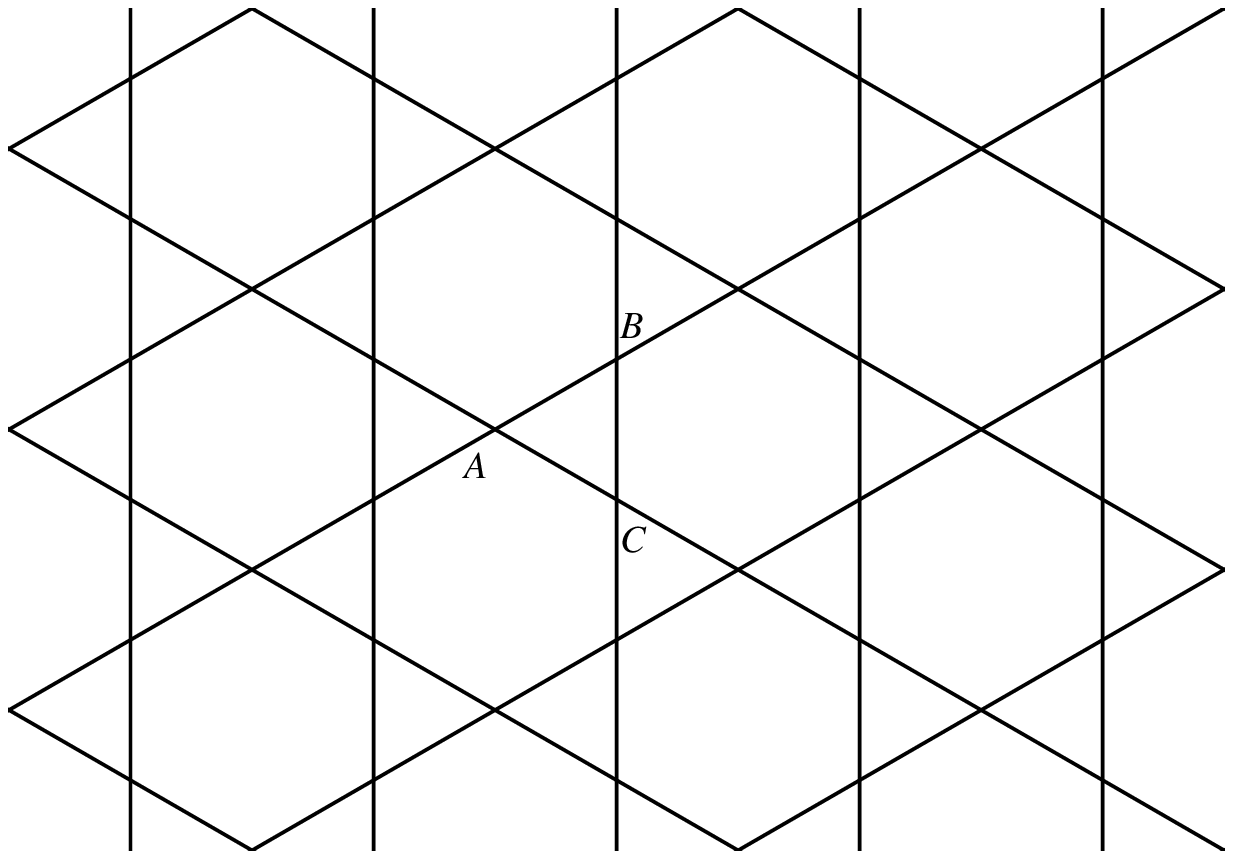,width=3 in} & 
\psfig{figure=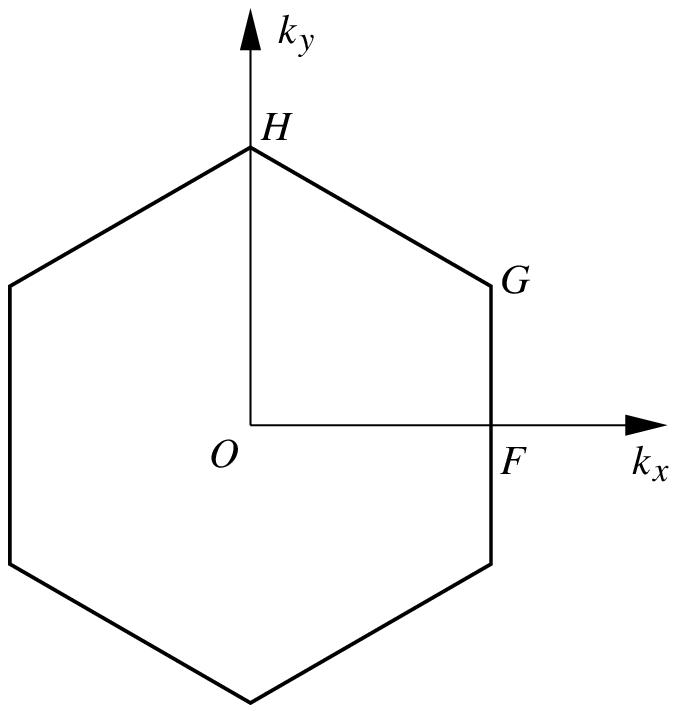,width=2.5 in}\\
(a) &(b)\\
\end{tabular}
}
\vspace{1 cm}
\caption{ (a) The \kag lattice is made up 
of triangular and hexagonal plaquettes 
with bond length $a$. 
Each unit cell has three sites, 
denoted here as $A(x,y)$, $B(x+\frac {\sqrt 3}2 a, y+\frac 12 a)$ and 
$C(x+\frac {\sqrt 3}2 a, y-\frac 12 a)$. 
(b) The first Brillouin zone in reciprocal space.  
Some points on the edge of the first Brillouin zone:
$F(\frac {\pi}{\sqrt 3 a},0)$, 
$G(\frac {\pi}{\sqrt 3 a},\frac {\pi}{3a})$, and  
$H(0, \frac {2\pi}{3a})$}
\label{fig:kaglattice}
\end{figure}



\begin{figure}
\centerline{
\begin{tabular}{cc}
\psfig{figure=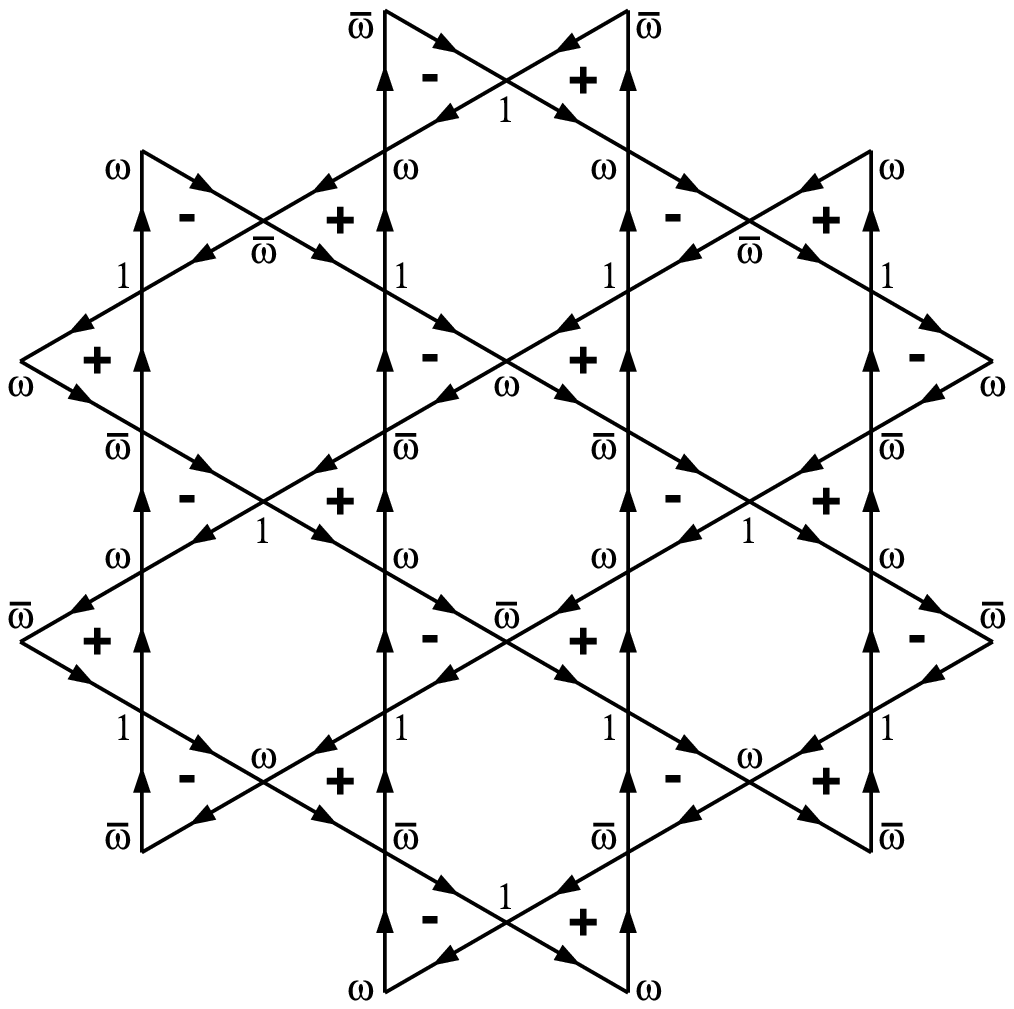,width=2.4 in} & \psfig{figure=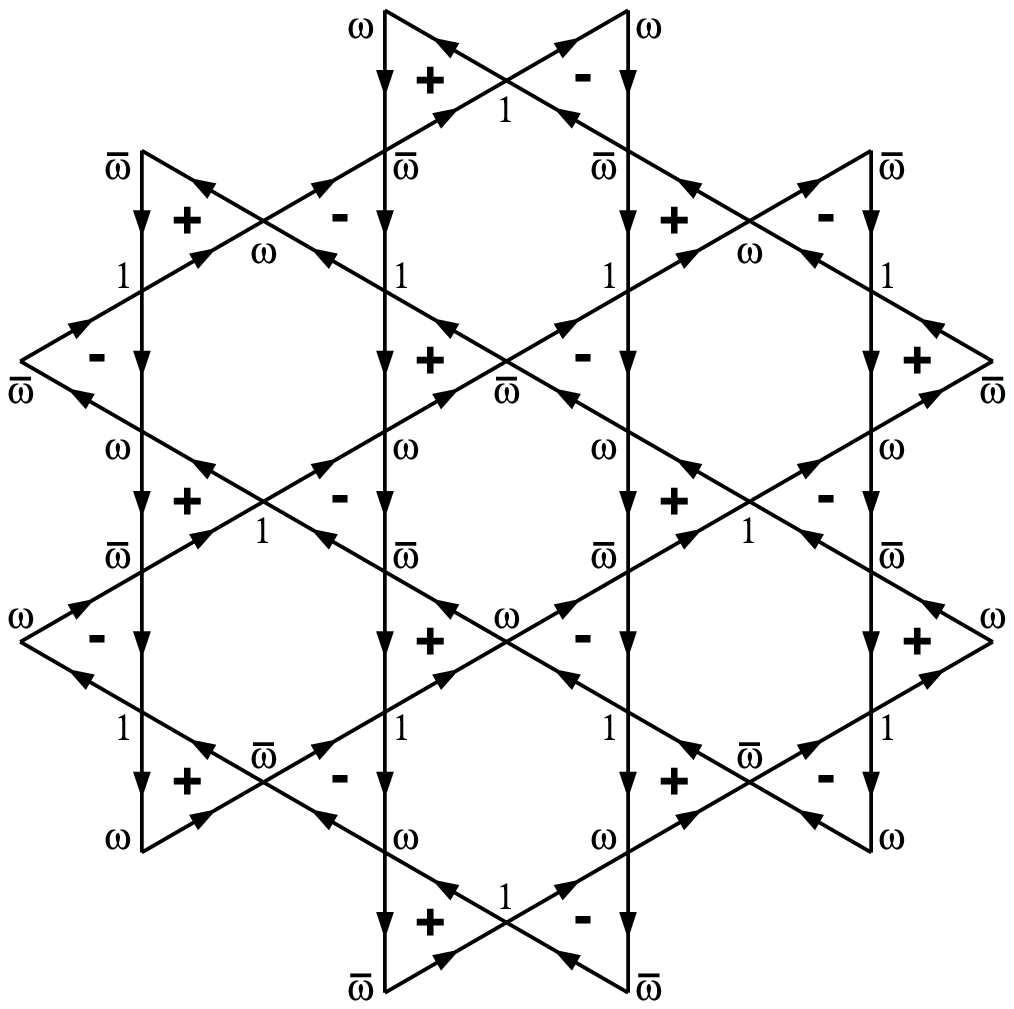,width=2.4 in}\\
(a) &(b)\\
\psfig{figure=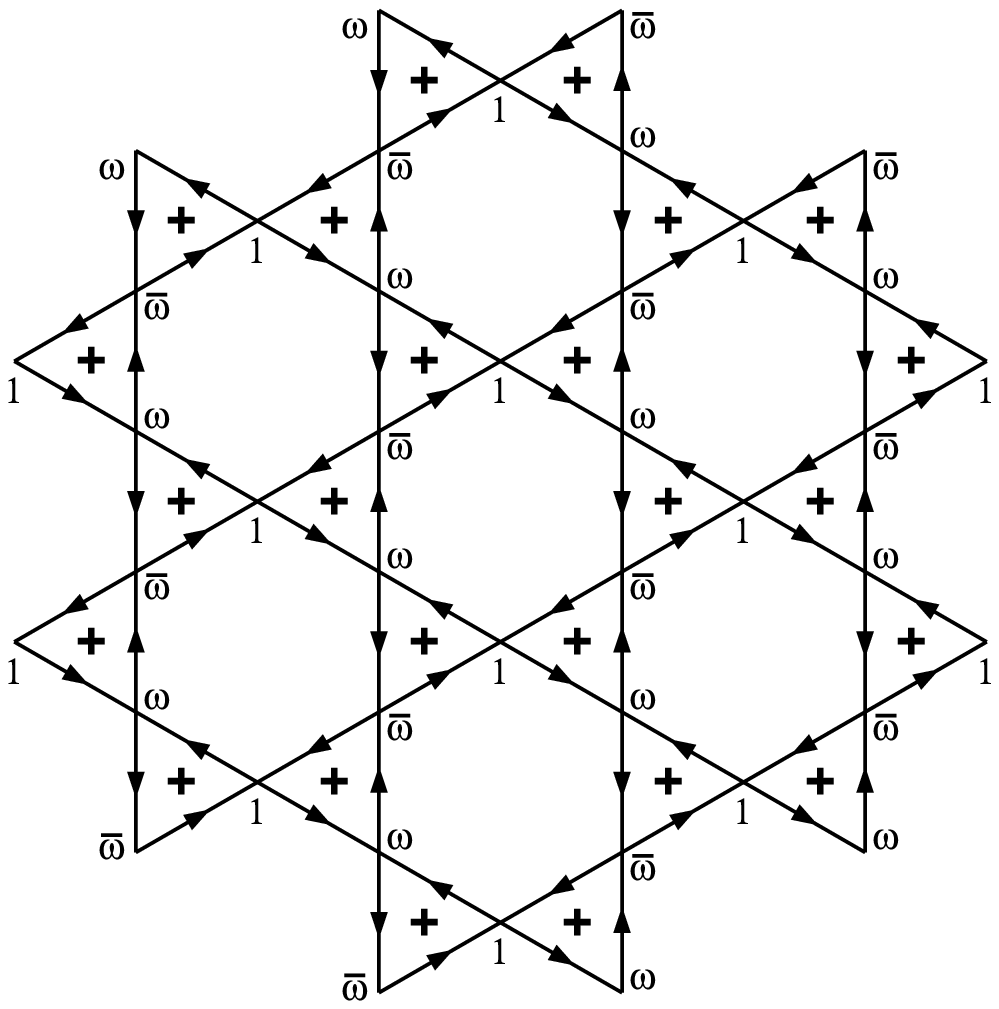,width=2.4 in} & \psfig{figure=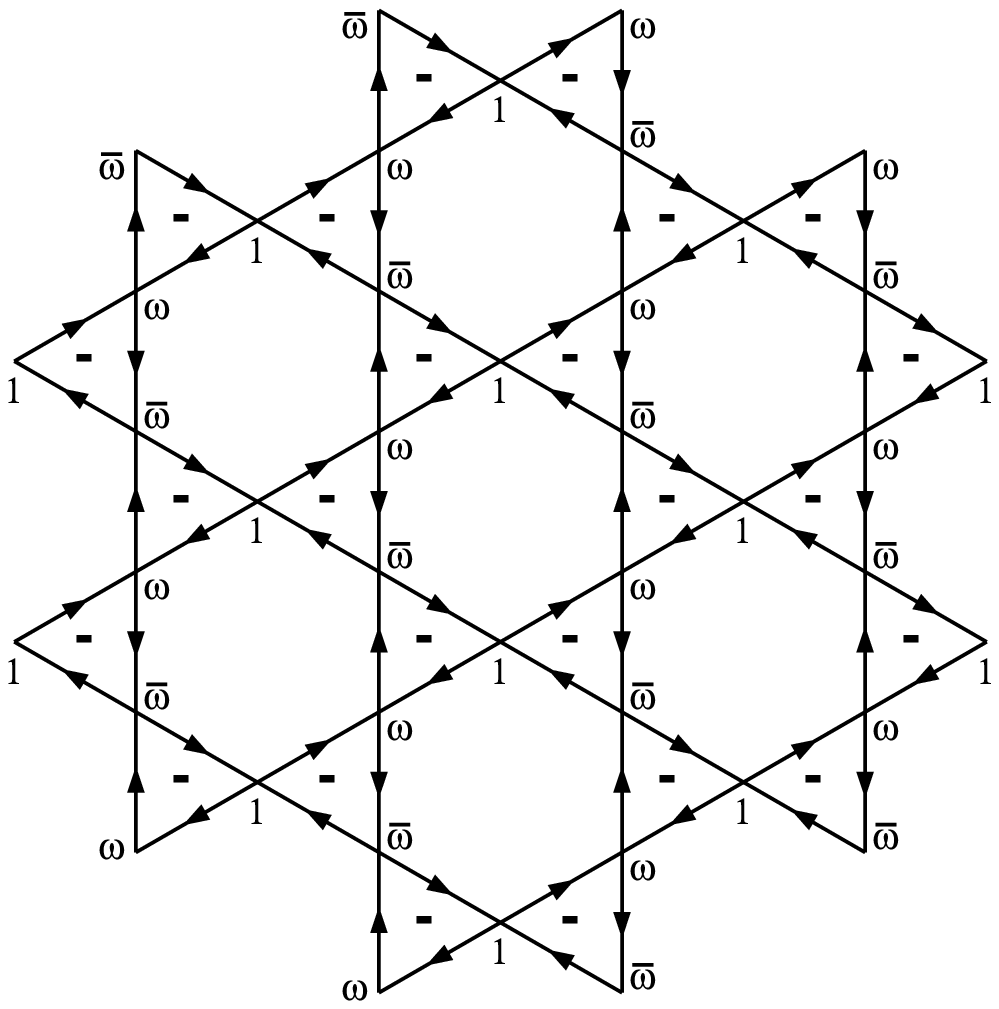,width=2.4 in}\\
(c) &(d)\\
\end{tabular}
}
\vspace{1 cm}
\caption{Some of the degenerate ground states with energy $\epsilon=-2$ 
in zero magnetic field. $\omega
=\exp(i\frac{2\pi}{3})$ and $\overline\omega$ is its complex conjugate.
The numbers next to the lattice points indicate the value of the wavefunction.
The arrows along the bonds indicate the currents (supercurrents in the
application to superconducting wire networks) and the signs at the centers
of the triangles indicate the sign of the circulation of the currents.
The $\vec k=0$ states are (c) and (d).}
\label{fig:kagomeMode}
\end{figure}


\begin{figure}
\centerline{
\psfig{figure=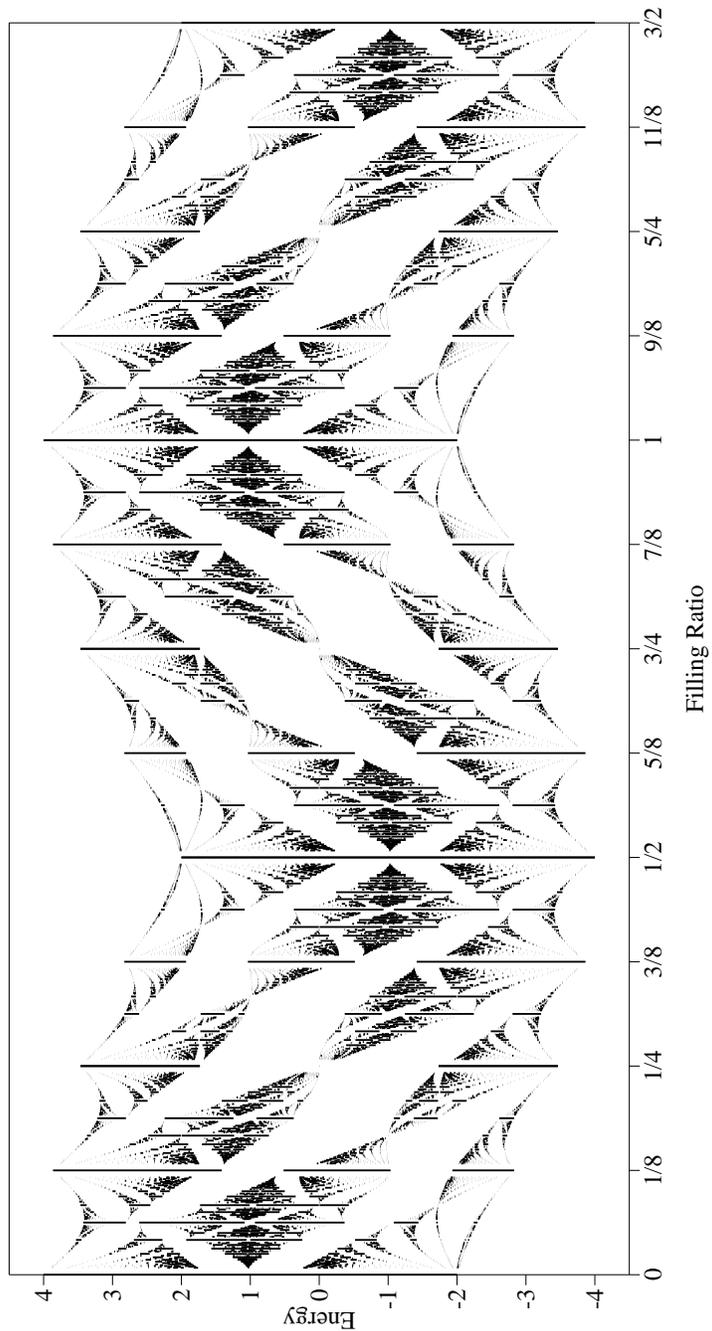,width=4 in}
}
\vspace{1 cm}
\caption{Spectrum of the \kag lattice tight-binding model in a 
uniform magnetic field.
Here and in the following figures the energy is in units of $t$.  
The filling ratio $f$ is the fraction of
a magnetic flux quantum passing through each elementary triangle of the 
\kag lattice.}
\label{fig:kagomespectrum0to1half}
\end{figure}

\begin{figure}[h!]
\centerline{
\psfig{figure=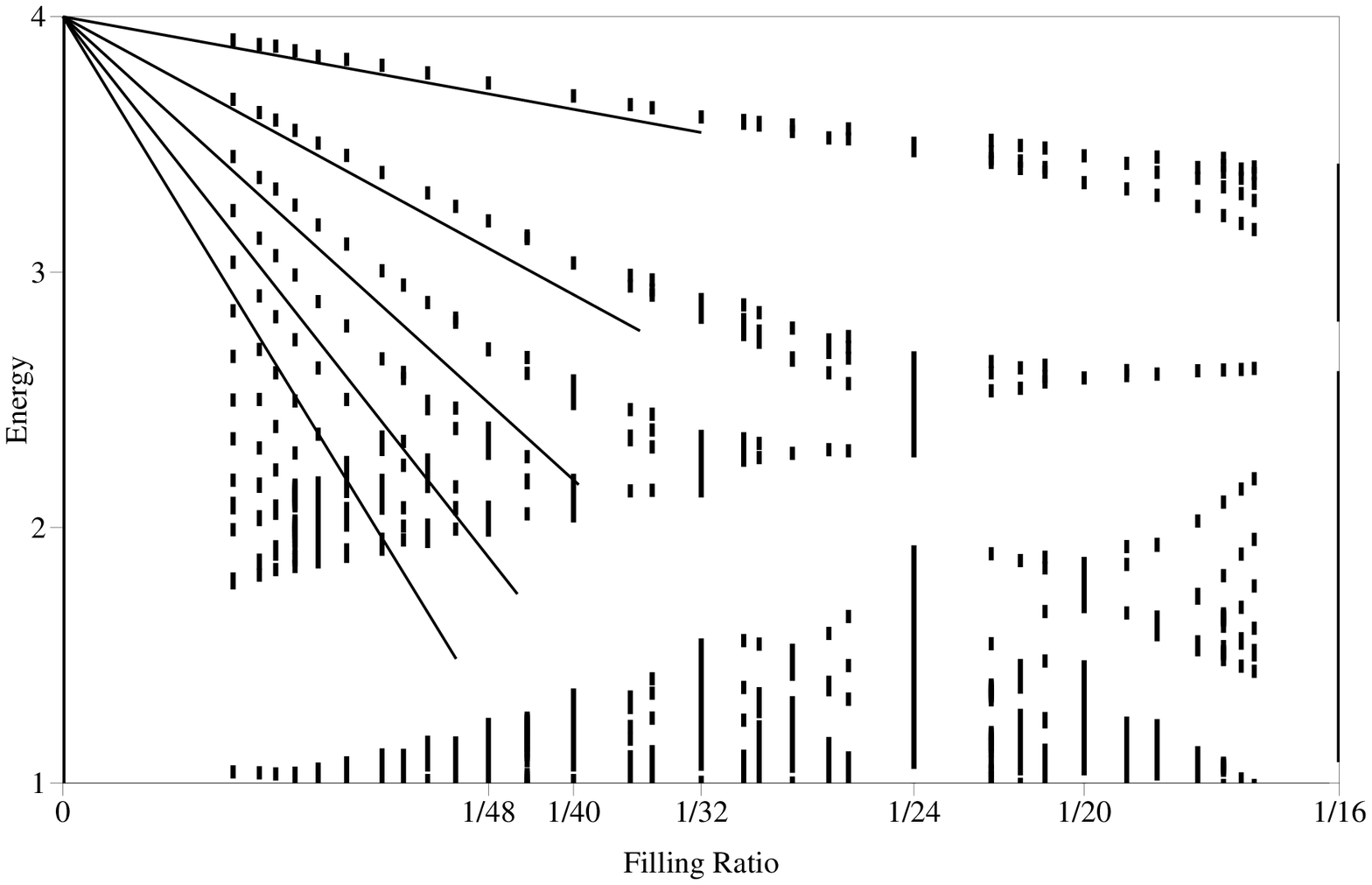,width=4 in}
}
\vspace{1 cm}
\caption{Energy spectrum for Landau levels near $\epsilon=4t$. The 
small $f$ behavior of the levels 
corresponding to
$n=0,1,2,3,4$ are indicated here as solid lines.}
\label{fig:UpperLandauLevel}
\end{figure}




\begin{figure}[h!]
\centerline{
\psfig{figure=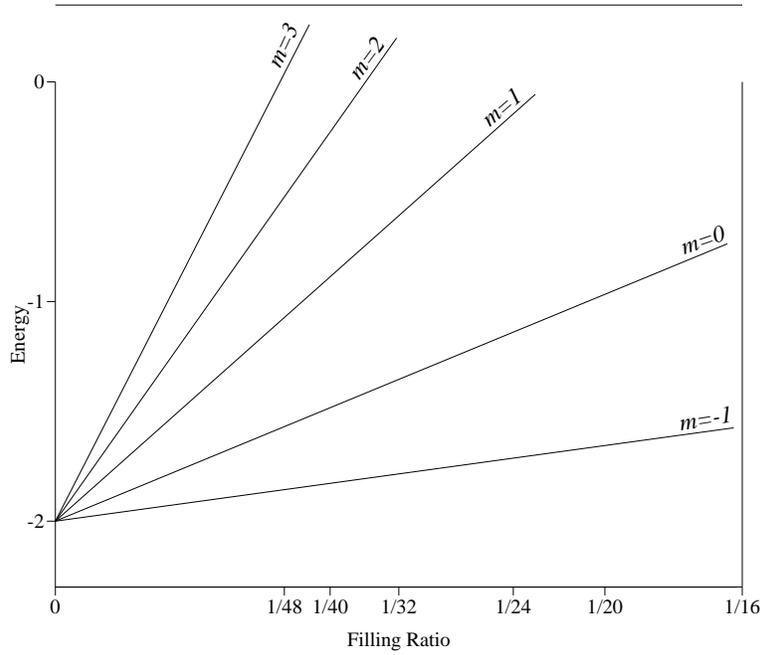,width=4in}
}
\vspace{1 cm}
\caption{Lower subband structure close to $\epsilon=-2t$.  The calculated
small $f$ behavior of some of the Landau levels are indicated by the
straight lines here.}
\label{fig:spectrumLowerLandau}
\end{figure}

\begin{figure}[h!]
\centerline{\psfig{figure=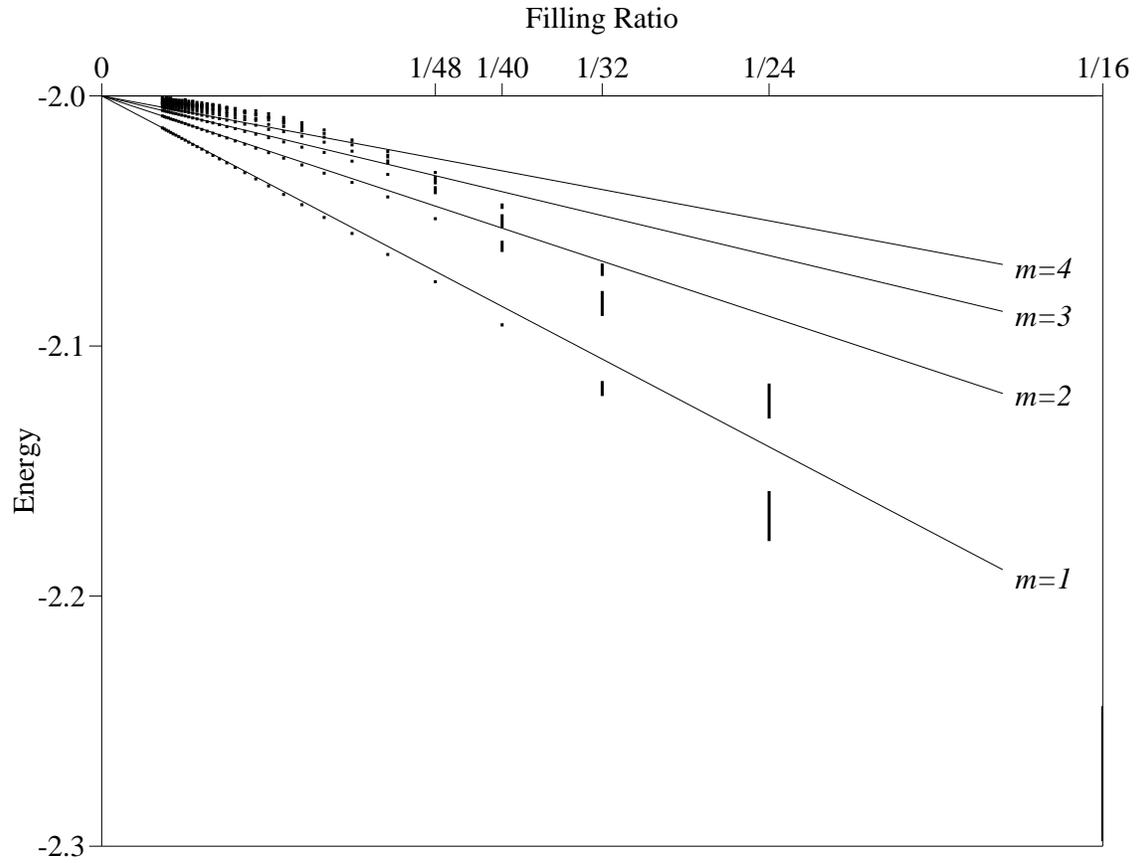,width=6 in}}
\vspace{1cm}
\caption{Subband structure below $\epsilon=-2t$ for filling ratios 
down to $f=1/256$.  The calculated small $f$ behavior of the 4 lowest
Landau levels are indicated by the straight lines.}
\label{fig:LowSpectrum}
\end{figure}

\begin{figure}
\centerline{
\psfig{figure=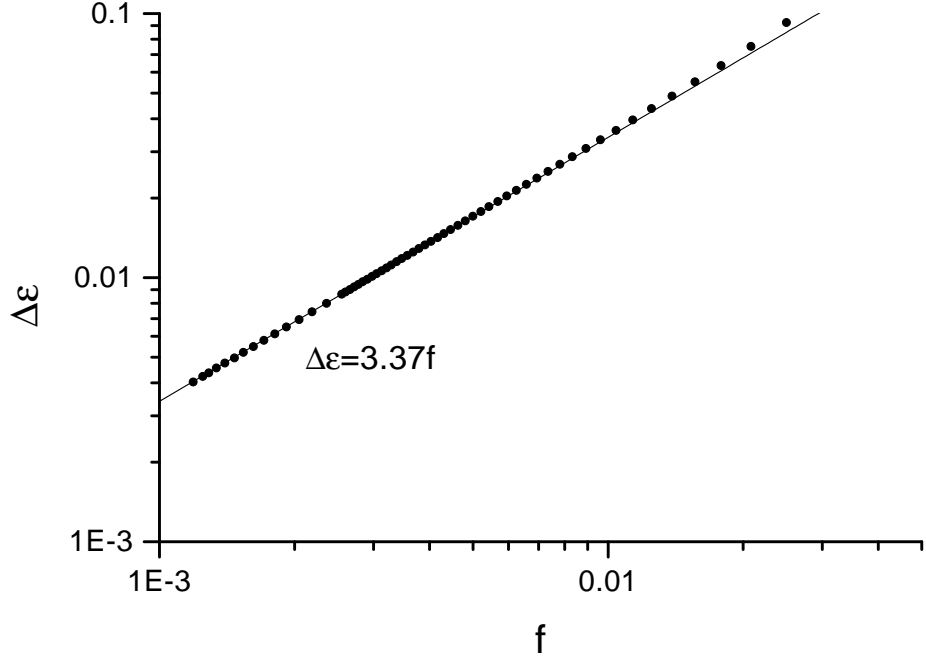,width=5 in}
}
\caption{Log-Log plot of the bottom edge of the spectrum close to 
$\epsilon=-2t$, going down to $f=1/840$, with 
$\Delta\epsilon=-2-(\epsilon_{min}/t)$.
The dots are from the 
numerics and the straight line indicates the expected small $f$ behavior.}
\label{fig:LogLowEdge}
\end{figure}

\begin{figure}
\centerline{
\psfig{figure=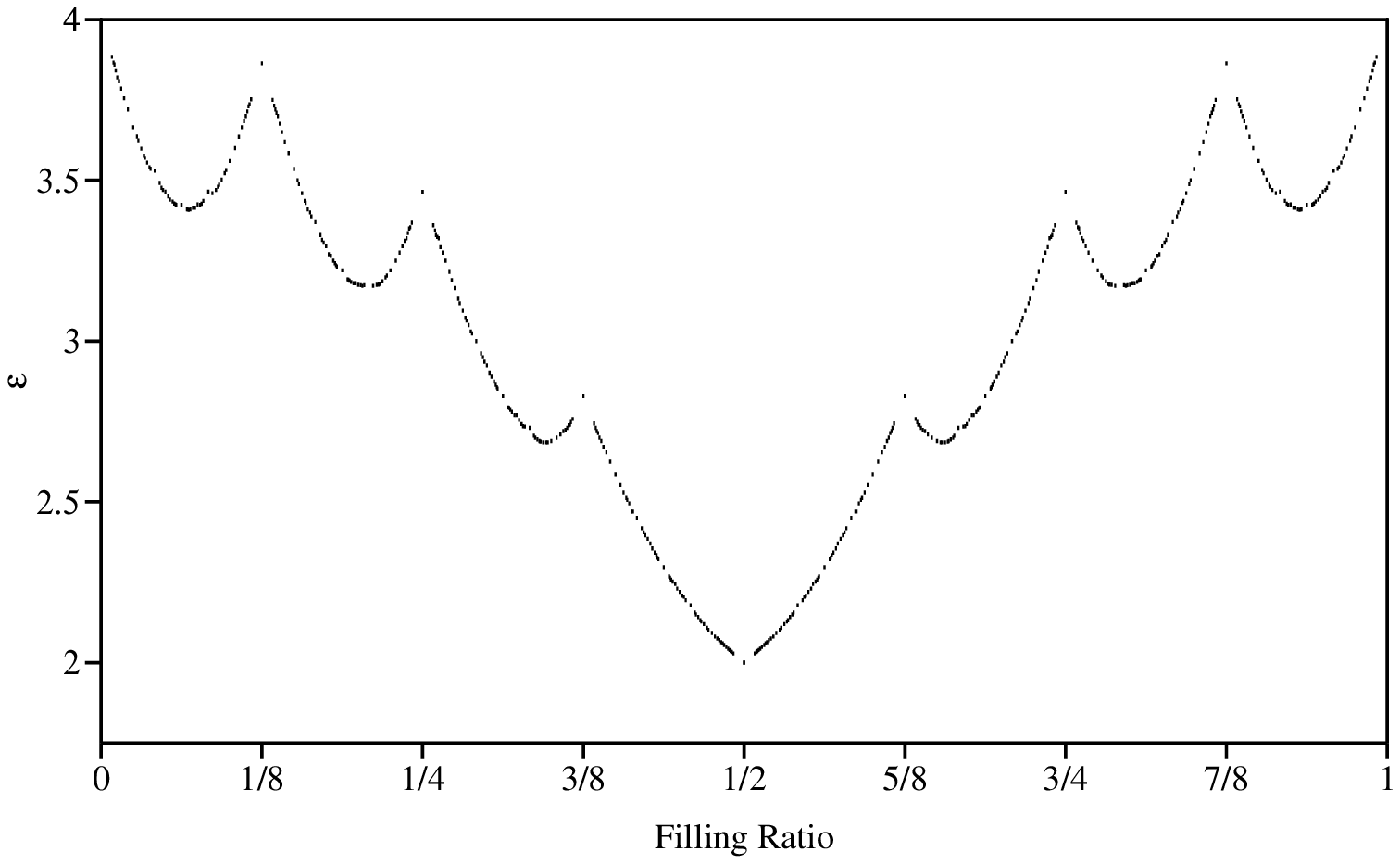,width=5 in}
}
\caption{Upper edge of the \kag lattice tight-binding
spectrum for $t>0$.  The mean-field transition temperature $T_c$
of a \kag wire-grid superconductor is linearly related to $\epsilon$,
so $T_c$ has a minimum at filling ratio $f=1/2$.}
\label{fig:upperedge}
\end{figure}

\end{center}

\begin{references}

\bibitem{Hofstadter76} D. R. Hofstadter,   Phys. Rev. B {\bf 14}, 2239 (1976).

\bibitem{Pannetier83} B. Pannetier, J. Chaussy, and R. Rammal, J. Phys. (Paris)
{\bf 44}, L853 (1983).

\bibitem{Mooij88} See, for instance, "Coherence in superconducting networks", edited by
J.E. Mooij and G.B.J. Sch\"{o}n, Physica {\bf 152B}, 1 (1988).

\bibitem{Alexander83} S. Alexander, Phys. Rev. B {\bf 27}, 1541 (1983).

\bibitem{Claro79} F. H. Claro and G. H. Wannier, Phys. Rev. B {\bf 19}, 6068 (1979).

\bibitem{Rammal85} R. Rammal, J. Physique {\bf 46}, 1345 (1985).

\bibitem{Vidal98} J. Vidal, R. Mosseri, and B. Dou{\c c}ot, 
Phys. Rev. Lett. {\bf 81}, 5888 (1998).

\bibitem{Pannetier01} B. Pannetier, C.C. Abilio, E. Serret, Th. Fournier,
P. Butaud and J. Vidal, Physica C {\bf 352}, 411 (2001).

\bibitem{Elser89} V. Elser, Phys. Rev. Lett. {\bf 62}, 2405 (1989).

\bibitem{Ritchey93} I. Ritchey, P. Chandra, and P. Coleman, 
Phys. Rev. B {\bf 47}, 15342 (1993).

\bibitem{Chalker92} J. T. Chalker, P. C. W. Holdsworth, and E. F. Shender, Phys. Rev. Lett. {\bf 68}, 
855 (1992).

\bibitem{HR92} D. A. Huse and A. D. Rutenberg, 
Phys. Rev. B {\bf 45}, 7536 (1992).

\bibitem{Reimers93} J. N. Reimers and A. J. Berlinsky, Phys. Rev. B {\bf 48}, 9539 (1993).

\bibitem{Villain80} J. Villian, R. Bidaux, J. P. Carton, and R. Conte, J. Phys. (Paris) {\bf 41}, 1263 (1980).

\bibitem{Greywall89} D. S. Greywall and P. A. Busch, Phys. Rev. Lett. {\bf 62}, 1868 (1989).

\bibitem{Greywall90} D. S. Greywall and P. A. Busch, Phys. Rev. Lett. {\bf 65}, 2788 (1990).

\bibitem{Ramirez91} For a brief review of this work, see A. P. Ramirez, J. Appl. Phys. {\bf 70}, 5952 (1991).


\bibitem{Higgins00}  M. J. Higgins, Y. Xiao, S. Bhattacharya, 
P.M.Chaikin, S.Sethuraman,
R. Bojko and D. Spencer, Phys. Rev. B {\bf 61}, R894 (2000); 
Y. Xiao, D. A. Huse, 
P. M. Chaikin, M. J. Higgins, S. Bhattacharya and D. Spencer, 
Phys. Rev. B {\bf 65}, 214503 (2002).

\bibitem{Lin94}  Y.-L. Lin and F. Nori, Phys. Rev. B {\bf 50}, 15953 (1994); Phys. Rev. B [in press].

\bibitem{Peierls33} R. E. Peierls, Z. Phys. {\bf 80}, 763 (1933).

\bibitem{yi} Yi Xiao, Princeton Univ. Physics Dept. Ph.D. thesis (2000).












%



\end{references}
\end{document}